\documentclass[a4paper,aps,prd,10pt,preprintnumbers,twocolumn,superscriptaddress,nofootinbib,amsmath,amssymb]{revtex4-1}
\usepackage{graphicx,cmap,hyperref}
\usepackage[utf8]{inputenc}
\usepackage[T1]{fontenc}

\begin{document}
\title{Can the abyss swallow gravitational waves or why we do not observe echoes?}
\author{Roman A. Konoplya} \email{roman.konoplya@gmail.com}
\affiliation{Research Centre for Theoretical Physics and Astrophysics, Institute of Physics, Silesian University in Opava, Bezručovo náměstí 13, CZ-74601 Opava, Czech Republic}
\author{Alexander Zhidenko} \email{olexandr.zhydenko@ufabc.edu.br}
\affiliation{Centro de Matemática, Computação e Cognição (CMCC), Universidade Federal do ABC (UFABC), \\ Rua Abolição, CEP: 09210-180, Santo André, SP, Brazil}
\begin{abstract}
Here we propose a simple explanation why echoes from wormholes mimicking black holes may be so small that they cannot be observed. The essence of the effect is in the redistribution of the initial energy of gravitational wave among multiple universes, connected by a wormhole.
\end{abstract}
\date{April 1, 2022}
\maketitle

\section{Introduction}

Recent years a hypothesis that a black hole might, in reality be, a wormhole constructed in such a special way, that it can mimic the black hole behavior, became extremely popular \cite{Cardoso:2016rao,Damour:2007ap}. Unless the motion of celestial bodies on the other side of the wormhole's throat significantly affects the matter in our spacetime \cite{Hong:2021qnk}, the observable proper oscillation frequencies called quasinormal modes \cite{Nollert:1999ji,Kokkotas:1999bd,Berti:2009kk,Konoplya:2011qq} as well as shadow and other optical characteristics would be the same for the black hole and the wormhole which is the black-hole mimicker. The only difference would come as a small modification of the gravitational quasinormal ringing at very later times, called echoes \cite{Abedi:2016hgu,Abedi:2017isz,Price:2017cjr,Wang:2018gin,Abedi:2018npz,Mannarelli:2018pjb,Barack:2018yly,Cardoso:2019apo,Bronnikov:2019sbx,Abedi:2020sgg,Dey:2020lhq,Chowdhury:2020rfj,Liu:2021aqh}. Therefore, the phenomenon of gravitational echoes attracted enormous attention of theoreticians and astrophysicists \cite{Cardoso:2016oxy,Foit:2016uxn,Holdom:2016nek,Barcelo:2017lnx,Mark:2017dnq,Cardoso:2017njb,Maselli:2017tfq,Cardoso:2017cqb,Bueno:2017hyj,Conklin:2017lwb,Wang:2018mlp,Correia:2018apm,Volkel:2018hwb,Du:2018cmp,Pani:2018flj,Tsang:2018uie,Testa:2018bzd,Mirbabayi:2018mdm,Oshita:2018fqu,Burgess:2018pmm,Sebastiani:2018ktb,Konoplya:2018yrp,Brustein:2018ixz,GalvezGhersi:2019lag,Barausse:2019pri,Wang:2019szm,Li:2019kwa,Wang:2019rcf,Conklin:2019fcs,Saraswat:2019npa,Maggio:2019zyv,Huang:2019veb,Hui:2019aox,Holdom:2019bdv,Fiziev:2019yjh,Churilova:2019cyt,Abedi:2020ujo,Oshita:2020dox,Oshita:2020abc,Holdom:2020onl,Kleihaus:2020qwo,Buoninfante:2020tfb,Fransen:2020prl,Maggio:2020jml,Dimitrov:2020txx,Liu:2020qia,Agullo:2020hxe,Singh:2020yau,Dey:2020wzm,Dey:2020pth,LongoMicchi:2020cwm,Vlachos:2021weq,Ikeda:2021uvc,Kirillov:2021bcs,Arimoto:2021cwc,Maggio:2021ans,Churilova:2021tgn,Yang:2021cvh,Zhang:2021fla,Ren:2021xbe,Fang:2021iyf,Annulli:2021gxr,Saraswat:2021ong,Roy:2021jjg,Ou:2021efv,Manikandan:2021lko,Chakravarti:2021clm,Coates:2021dlg,Huang:2021qwe,Bao:2022vtf,Wu:2022eiv,Mukherjee:2022wws,Lin:2022azn,Ma:2022xmp}.
At the same time, numerous attempts to observe echoes have not succeeded so far \cite{Ashton:2016xff,Westerweck:2017hus,Nielsen:2018lkf,Lo:2018sep,Uchikata:2019frs,Tsang:2019zra,Datta:2019epe,LIGOScientific:2020tif,LIGOScientific:2021sio}. Here we suggest a simple explanation why we might not be able to observe gravitational echoes.

In \cite{Cardoso:2016rao} Cardoso, Franzin, and Pani suggested the model of a wormhole which consists of two Schwarzschild solutions in two universes or two separated asymptotic regions of the same universe
\begin{eqnarray}\label{CFPwormhole}
  ds^2&=&-g(r)dt^2+\frac{dr^2}{g(r)}+r^2(d\theta^2+\sin^2\theta d\phi^2), \\\nonumber g(r)&=&1-\frac{2M}{r},
\end{eqnarray}
identified at the throat whose radius is larger than the Schwarzschild radius, $r_0>2M$.

\begin{figure}
\includegraphics[width=\linewidth]{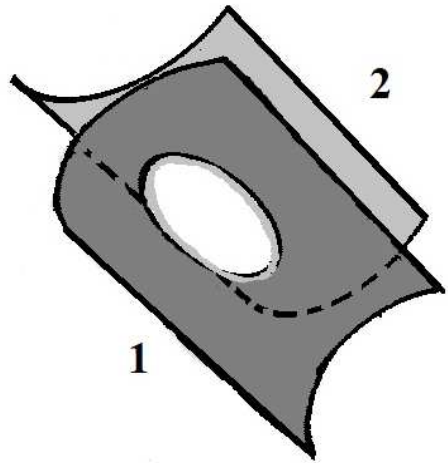}
\caption{Simple wormhole connecting two universes.}\label{2universes}
\end{figure}

Another approach, proposed by Damour and Solodukhin \cite{Damour:2007ap}, can be described by the following line element:
\begin{equation}\label{DSwormhole}
  ds^2=-\left(g(r)+\lambda^2\right)dt^2+\frac{dr^2}{g(r)}+r^2(d\theta^2+\sin^2\theta d\phi^2),
\end{equation}
for which the identification occurs at the radius of the throat $r_0=2M$ connecting the two universes or separated regions. In this case the asymptotic mass is
$$\frac{M}{1+\lambda^2},$$
being again smaller than half of the wormhole radius.
Both models lead to the same topological configuration (see Fig.~\ref{2universes}).

\section{Wormholes connecting multiple universes}

\begin{figure}
\includegraphics[width=\linewidth]{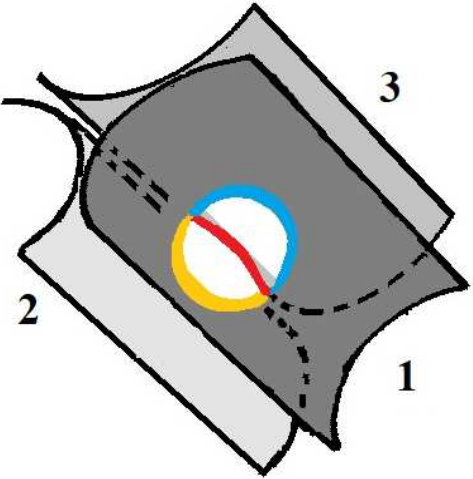}
\caption{A wormhole connecting three universes.}\label{3universes}
\end{figure}

We suggest a straightforward generalization of the above wormhole configurations. First, we will consider a simpler model and assume that a wormhole connects three universes or separated regions, that is, it consists of a ``tunnel'' in ``our'' Universe 1, which further splits into the two tunnels leading to Universe 2 and Universe 3. When we neglect the size of the tunnels, we describe the geometry by either (\ref{CFPwormhole}) or (\ref{DSwormhole}) metric tensor, but identify each half of the surface of the sphere $r=r_0$ with different Universes (see Fig.~\ref{3universes}).
\begin{enumerate}
\item We identify the points of the Universe 1 and Universe 2 on the sphere $r=r_0$, once $0\leq\phi\leq\pi$ (yellow).
\item If $\pi\leq\phi\leq2\pi$ we identify the points of the Universe 1 and Universe 3 on the same sphere $r=r_0$ (blue).
\item We also identify the remaining halves of the spheres  $r=r_0$ of the Universes 2 and 3 (red).
\end{enumerate}
As a result of the above identifications, the circumference
$$r=r_0, \quad 0\leq\theta\leq\pi, \quad \phi=0,\pi$$
belongs to all three universes.

\begin{figure}
\includegraphics[width=\linewidth]{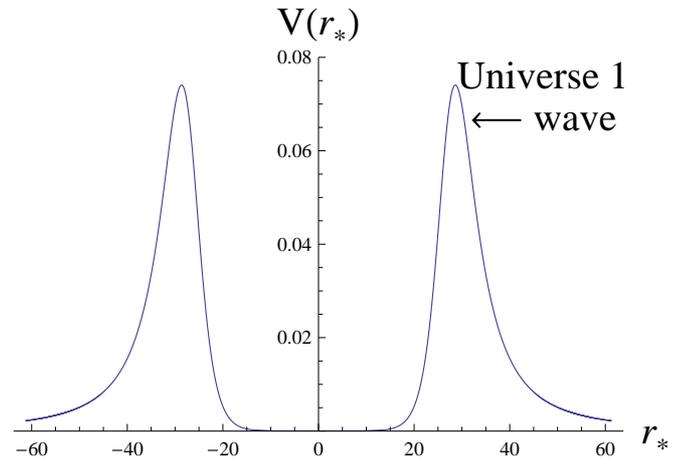}
\caption{Effective potential for the wormhole perturbations.}\label{potential}
\end{figure}

Let us recall the gravitational-wave propagation once a wormhole, which connects two universes, is perturbed. The perturbation equations can be reduced to a wave-like equations for each multipole numbers, $\ell$ and $m$,
\begin{equation}\label{wavelike}
  \frac{\partial^2\Psi}{\partial t^2}-\frac{\partial^2\Psi}{\partial r_*^2}+V(r_*)\Psi=0,
\end{equation}
with an effective potential $V(r_*)$, which consists of two potential barriers, one in each universe (see Fig.~\ref{potential}). Starting its way in the first universe, the signal is scattered by the potential barrier. A part of the wave, which is reflected by the barrier in the first universe, can be seen by a distant observer as quasinormal ringing of the wormhole. However, the other part of the wave which passes through the throat is later scattered again by the symmetric barrier in the other universe (see Fig.~\ref{potential}). The wave reflected by the symmetric barrier in the other universe propagates back through the throat, and, after tunneling to the remote observer in the Universe 1, can be observed as another pulse of the quasinormal ringing called echoes.

Now let us discuss propagation of gravitational waves when the wormhole, connecting three universes, is perturbed. Since, for the sake of simplicity, we consider the model, which is described by the same line element (and only different identifications at the throat), the effective potential barrier remains the same. However, now there is an important difference -- at the throat only half of the signal goes to the Universe 2, while the other half goes to the Universe 3.

Then, the initial quasinormal ringing phase as observed in the first universe remains the same, as it would be for a simple wormhole connecting two universes, and mimics the signal from the Schwarzschild black hole. However the parts of the wave which passed the throat and experienced secondary scattering in the other two universes will be changed.
\begin{enumerate}
\item After propagating back through the throat the reflected wave loses approximately half of its energy, which goes from the Universe 2 to the Universe 3 and from the Universe 3 to Universe 2 instead of returning to the universe 1. Thus, the echoes observed in the universe 1 come with much smaller energy (amplitude).
\item Due to symmetry breaking at the throat, the signal in Universe 2 or Universe 3 is not simply half of the ingoing signal -- it corresponds to the wave, which passes through a half of the spherical throat. It means that multipolar decomposition is mixed, and, for instance, the $\ell=m=2$ wave contributes to all the multipoles on the other side of the throat. In other words, there is redistribution of energy among various multipole numbers. On its way back through the barrier the same phenomenon occurs, further dissipating the energy of $\ell=m=2$ echoes into various harmonics, which have much smaller initial amplitude. Similarly, another dominant mode $\ell=m=3$ dissipates to all the multipoles twice, while passing the wormhole throat back and forth.
\end{enumerate}
Thus, because of the above two phenomena, the splitting of energy between universes and redistribution of energy among the harmonics, which cannot be practically detected, in our model the echoes are much weaker as compared to the case of a wormhole connecting only two universes.

If we consider a wormhole, which connects more than three universes, it is clear that the resulting echoes will be even much smaller, because more of its energy dissipates to other universes. Indeed, if we identify $$r=r_0, \qquad (k-2)\frac{2\pi}{n}\leq\phi\leq(k-1)\frac{2\pi}{n}$$ in the Universe 1 with the corresponding points in the Universe $k=2,3,\ldots,(n+1)$, then only $1/n$ of the echo energy goes into the Universe 1, while most part of it is redistributed among other $n$ universes.

\section{Conclusion}
We conclude that echoes can be unobservable when the wormhole, which mimics a black hole, connects not two, but many universes. In order to illustrate this conclusion we have considered a simple model. However, our conclusions remain the same in more complex configurations, such as, for instance, a connection between the universes through the Slow zone (the Hub) proposed by the James S. A. Corey in \href{https://en.wikipedia.org/wiki/The_Expanse_(novel_series)}{series of novels ``The Expanse''}.

\begin{acknowledgments}
We acknowledge Barbara Idino Konoplya for suggesting this idea to us.
A.~Z. was supported by Conselho Nacional de Desenvolvimento Científico e Tecnológico (CNPq).

\end{acknowledgments}

\end{document}